\documentclass[draftclsnofoot,onecolumn]{IEEEtran}
\makeatletter
\def\ps@headings{%
\def\@oddhead{\mbox{}\scriptsize\rightmark \hfil \thepage}%
\def\@evenhead{\scriptsize\thepage \hfil \leftmark\mbox{}}%
\def\@oddfoot{}%
\def\@evenfoot{}}
\makeatother
\usepackage{multicol}
\usepackage{cite}
\usepackage[dvips]{graphicx}
\usepackage{epsfig}
\graphicspath{{eps/}}
\usepackage{subfigure}
\usepackage{float} 
\usepackage{gensymb}
\usepackage{array}
\usepackage{algorithm}
\usepackage{algorithmic}
\usepackage{mdwmath}
\usepackage{mdwtab}
\usepackage{url}
\begin{document}
\title{CogCell: Cognitive Interplay between 60\,GHz Picocells and 2.4/5\,GHz Hotspots in the 5G Era}
\author {\IEEEauthorblockN{Kishor Chandra, R. Venkatesha Prasad, Bien Quang, I.G.M.M. Niemegeers}\\
\IEEEauthorblockA{EEMCS Faculty, Delft University of Technology, The Netherlands\\
%Mekelweg 4, 2628 CD, Delft, The Netherlands \\
Email: \{k.chandra, v.q.bien, i.g.m.m.niemegeers\}@tudelft.nl, rvprasad@ieee.org.}
}
\maketitle
\begin{abstract}
Rapid proliferation of wireless communication devices and the emergence of a variety of new applications have triggered  investigations into next-generation mobile broadband systems, i.e., 5G. Legacy 2G--4G systems covering large areas were envisioned to serve both  indoor and outdoor environments. However, in the 5G-era, 80\% of overall traffic is expected to be generated in indoors. Hence, the current approach of macro-cell mobile network, where there is no differentiation between indoors and outdoors, needs to be reconsidered. We envision 60\,GHz mmWave picocell architecture to support high-speed indoor and hotspot communications. We envisage the 5G indoor network as a combination of-, and interplay between, 2.4/5\,GHz having robust coverage and 60\,GHz links offering high datarate. This requires an intelligent coordination and cooperation. We propose 60\,GHz picocellular network architecture, called CogCell, leveraging the ubiquitous WiFi. We propose to use 60\,GHz for the data plane and 2.4/5GHz for the control plane. The hybrid network architecture considers an opportunistic fall-back to 2.4/5\,GHz in case of poor connectivity in the 60\,GHz domain. Further, to avoid the frequent re-beamforming in 60\,GHz directional links  due to mobility, we propose a cognitive module -- a sensor-assisted intelligent beam switching procedure -- which reduces the communication overhead. We believe that the CogCell concept will help future indoor communications and possibly outdoor hotspots, where mobile stations and access points collaborate with each other to improve the user experience.
\end{abstract}
\IEEEpeerreviewmaketitle
\section{Introduction}
The unprecedented but anticipated massive growth of mobile data traffic is posing many challenges for 5G communication systems. 5G networks aim to achieve ubiquitous communication between anybody and anything, anywhere and at anytime. The performance requirements are far beyond what is offered by current systems -- in particular a 1000x increase in network capacity is targeted. All this requires new network architecture and technologies. Moreover new spectrum will be needed. For example, millimeter wave (mmWave) communication requires very different approaches for PHY, MAC and network layers. The general consensus among researchers and industry is that 5G will not be a mere incremental evolution of 4G~\cite{Jefry5G}. However, 2G -- 4G will have to be integrated with the new technologies to ensure the support of legacy systems.
 
\begin{figure*}[t]
	\centering
	\includegraphics[width=6.0in,height=3.4in]{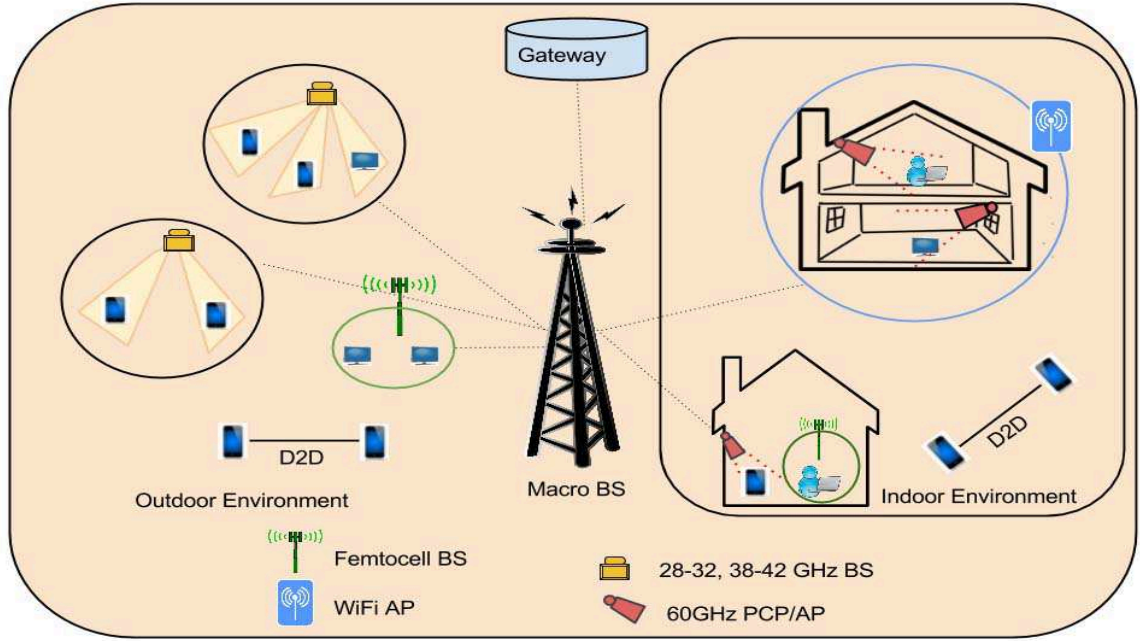}
		\hspace{5cm}\caption{A 5G scenario with multiple radio access technologies.}
	\label{5Gscenario}
	\end{figure*}

Fig.~\ref{5Gscenario} shows 5G communication scenario, where multiple radio access technologies (RATs), i.e., 60\,GHz Wireless Local Area Networks (WLAN), 2.4/5\,GHz WiFi, 28-30 or 38-40\,GHz outdoor mmWave base stations (BSs) and macro \& femto cell  BSs are present.  For efficient spectrum utilization, multiple licensed as well as unlicensed bands will need to work in cohesion for different applications. mmWave based mobile communication (28-32 and 38-42\,GHz spectrum) and WLANs at 60\,GHz will coexist with  legacy cellular networks and WLANs. Thus 5G spectrum would span from sub-GHz to mmWave frequency bands to support diverse applications and services. To exploit the available spectrum across the various frequency bands,  a highly flexible communication interface is required which can support multiple RATs for various, possibly very different, services  at the same time. To meet the above stated requirements, various solutions are being discussed. We summarize them as follows.

\textit{Network architecture}: Instead of a rigid and infrastructure-centric approach adopted by previous generations, device- and user-centric architectures are being advocated for 5G, in order to better support ubiquitous and seamless communication. Further, the concept of cloud-based radio access network (C-RAN) is proposed to reduce operational costs by efficient utilization of radio resources~\cite{chinamobile}. In C-RAN, traditional base station functionality such as baseband processing and resource allocation is offloaded to a central location, to provide dynamic resource allocation leading to a better utilization of baseband processing resources.  Another architectural change expected is the macro-assisted small cells -- also called \textit{phantom cells}~\cite{phantomcells}. In this approach, the control plane and data plane are decoupled. The macro cell covering a large area is responsible for the control and management functions, while small cells are used solely for providing high datarate communications. Usually small cells remain in a turn-off state to save energy. Furthermore, for devices which are in the proximity of each other, direct device-to-device (D2D) communication is considered and is expected to become an integral part of 5G.
	
\textit{Medium access control and signaling}:
5G need to support a variety of applications, which are very different in terms of traffic patterns, datarates and latency constraints. For example, machine-to-machine (M2M) communication will have infrequent small packets with low datarates but with critical latency requirements. Video applications, e.g., 4K video, have some latency requirements, can tolerate errors to an extent, but will require very high datarates. Web browsing and file sharing applications, on the other hand have again different requirements. In case of M2M, signaling and control mechanisms employed in current networks would cause high overheads. Widespread use of M2M may lead to situations where thousands of devices try to access a channel simultaneously. Current access mechanisms are not designed to do this. Furthermore, to enable D2D, very efficient signaling mechanisms are required so that spectrum utilization can be increased. In case of ultra-dense networks, coordination among small cells, needed to mitigate the interference, will lead to high signaling overhead. Thus flexible medium access and signaling protocols are needed to optimize the channel utilization for a wide variety of applications  

\textit{Physical layer techniques}: From the perspective of the physical layer, to combat the scarcity of available radio spectrum in the lower frequency bands, mmWave frequencies (30\,GHz to 300\,GHz) are being explored as alternatives for both outdoor and indoor communication due to the huge bandwidth they provide. Licensed 28-30\,GHz and 38-42\,GHz bands are suitable for outdoor cellular networks~\cite{mmwaveMobilenetworks}, while the unlicensed 60\,GHz band is suitable for indoor communication due to its propagation characteristics~\cite{mmwaveWLAN}.  Another breakthrough technology which will certainly have a distinct place in 5G is Massive MIMO~\cite{mmwavebeamforming, massiveMIMO}. In Massive MIMO the number of antennas at a BS  are much higher than the number of devices being served. This enables simple \textit{spatial multiplexing} and \textit{demultiplexing}. The small size of antennas and antenna spacing at mmWave frequencies make massive MIMO a suitable beamforming technology for devices\footnote{We use the term device to mean a mobile station or a handheld user equipment.} as well as BSs.

It is predicted that by year 2020 indoor/hotspot traffic will account for 80-90$\%$ of total traffic volume~\cite{cisco}. Datarates on the order of multi-Gb/s will be required in indoor environment to support high definition video streaming and gaming applications.  Existing 3G and 4G systems were designed to support the same set of services both in indoor and outdoor environments. However, this will not be the case in 5G. A variety of services are emerging and many of them, in particular, high datarate uncompressed video will be mainly confined to indoors and hotspots. 
Therefore, 5G networks must take care of the traffic dissimilarities between indoor and outdoor environment. To tackle this challenge, high capacity indoor local small cells need to be designed that can provide multi-Gb/s connectivity with better coverage.

The 60\,GHz frequency band has emerged as the most promising candidate for high speed indoor communications. However, its inability to penetrate walls poses a serious challenge for providing seamless connectivity. Further, the use of narrow beamforming makes it challenging to support mobile devices, due to the link outages caused by antenna beam misalignment resulting from mobility of users. This requires beam tracking and adaptive beamforming. We propose CogCell concept, a 2.4\,GHz assisted 60\,GHz picocellular network architecture in which 60\,GHz is used for high speed data communication (data-plane traffic) while 2.4/5 GHz WiFi is used for control purposes (control-plane traffic). Several 60\,GHz picocells are managed by a single WiFi cell thus facilitating easy and robust network and mobility management with picocells. In the absence of a 60\,GHz  link, 2.4\,GHz can also be used as a fall-back data-plane option in CogCell making the best of both worlds.  The problem of frequent re-beamforming in 60\,GHz can be circumvented by leveraging the sensing and processing capabilities of smart devices that are using the 60\,GHz links. We will show how motion sensors (present in smart phones and tablets) will be used to predict user movement and thus maintain the beam alignment. CogCell architecture has many features: (i)~Better spectrum utilization by switching between 2.4 and 60\,GHz bands for control and data transmission, respectively; (ii)~Opportunistic fall-back to 2.4\,GHz band for data transmission, if the 60\,GHz link is not available; and (iii)~Sensor assisted cognitive and adaptive beam-tracking which reduces the need for frequent re-beamforming of 60\,GHz links in case user devices move. 

\section{60\,GHz Communication for Multi-Gb/s Indoor connectivity}
\label{Issues_802.11ad}
Despite very sophisticated PHY/MAC layer techniques such as MU-MIMO, higher order modulations, channel bonding and frame aggregation, it is hard to improve the WiFi datarate further. For example, despite using channel bonding and multi-user MIMO schemes IEEE 802.11ac can only provide a peak datarate of around 1\,Gb/s because of limited bandwidth in the 2.4/ 5\,GHz frequency bands. On the other hand, large bandwidth is available in unlicensed 60\,GHz band. The 60\,GHz MAC standards IEEE 802.15.3c~\cite{iee:IEEE802.15.3c} and IEEE 802.11ad~\cite{IEEE802.11ad} has already been completed, providing datarates up to 5-7\,Gb/s  for a range of 10 to 20\,m. IEEE 802.11ad is backward compatible with IEEE 802.11b/g/n/ac. However, there remain several issues which need to be addressed to realize multi-Gb/s 60\,GHz indoor networks. 
\begin{quote}
\textbf{Access delay}: 60\,GHz devices and access points (AP) employ directional antennas to compensate for free space path loss. IEEE 802.11ad and IEEE 802.15.3c divide the area around an AP in sectors, e.g., a sector can span over 60\degree or 90\degree. CSMA/CA based random access is used during predefined time periods -- in each sector in a round robin fashion -- called Contention Based Access Period (CBAP). A device has to wait for the CBAP period allocated to its sector. For example, if each sector spans an angle of 90\degree then there are four sectors. Thus, if a device generates a request just after the allocated CBAP period for its sector, it has to wait until the next three CBAP periods. This could introduce a considerable amount of delay before the request is fulfilled. 

\textbf{Re-beamforming}: Although the peak PHY datarate promised by IEEE 802.11ad is about 7\,Gb/s, realizing a seamless multi-Gb/s WLAN system providing a sustained peak datarate is difficult. 60\,GHz links are highly susceptible to blockage caused by obstacles such as humans, furniture, walls, etc. Further, communication using narrow beams has to track moving devices to maintain the link. With narrow beams, beam misalignment caused by small movements may result in broken links. If a device moves away from the beam coverage area, an exhaustive beam-search is required, resulting in excessive delays and communication overhead. It is therefore important to keep beam alignment in order to maintain a stable link.

\textbf{Hand-off}: While using directional antennas at 60\,GHz, AP/device discovery and fast handover are difficult. Since 60\,GHz signals cannot penetrate walls, there will be many  60\,GHz APs in an indoor area. This can result in frequent hand-off when a user moves in the indoor area. When moving from one room to another, one should be able to quickly reconnect with another AP. To ensure this, fast discovery and authentication are needed. Since the datarate is very high, a small interruption in signal coverage can lead to the loss of a large amount of data. Further, frequent device discovery and association could lead to excessive energy consumption resulting in fast battery drain. 
\end{quote}

To address the above issues, we propose to use WiFi and mmWave CogCell hybrid architecture. This will enable smooth network management, fast channel access and device discovery. Here WiFi supports control plane functions while 60\,GHz offers data plane functionality.  To avoid frequent re-beamforming caused by mobility, we employ motion sensors to predict the next location of the user so that appropriate beam switching can be performed. 
\section{Indoor networks based on combination of WiFi and 60\,GHz communication}\label{hybridarch}
In this section we discuss the capacity and coverage limitations of 2.4/5 and 60\,GHz signals, respectively. We illustrate that 2.4\,GHz and 60\,GHz systems are complementary in terms of coverage and capacity, and explain how the proposed CogCell architecture enables interplay of both to provide a robust multi-Gb/s WLAN connectivity. 

\subsection{Complementarity of 2.4 and 60 GHz}\label{complementary}
\begin{figure}
\centering
\subfigure[Signal coverage comparison at 2.4 GHz and 60 GHz.]{ 
%\centering
\includegraphics[width=6.0in,height=2.4in]{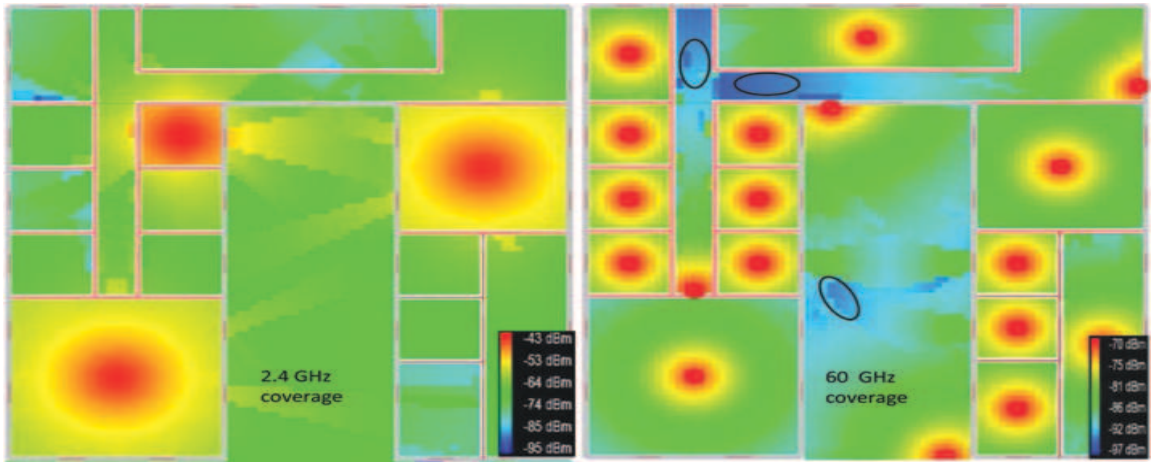}
\label{fig:coverage24}
}
\subfigure[Comparison of peak datarates at 2.4/5\,GHz and 60\,GHz.]{
\includegraphics[width=4.0in,height=4.4in]{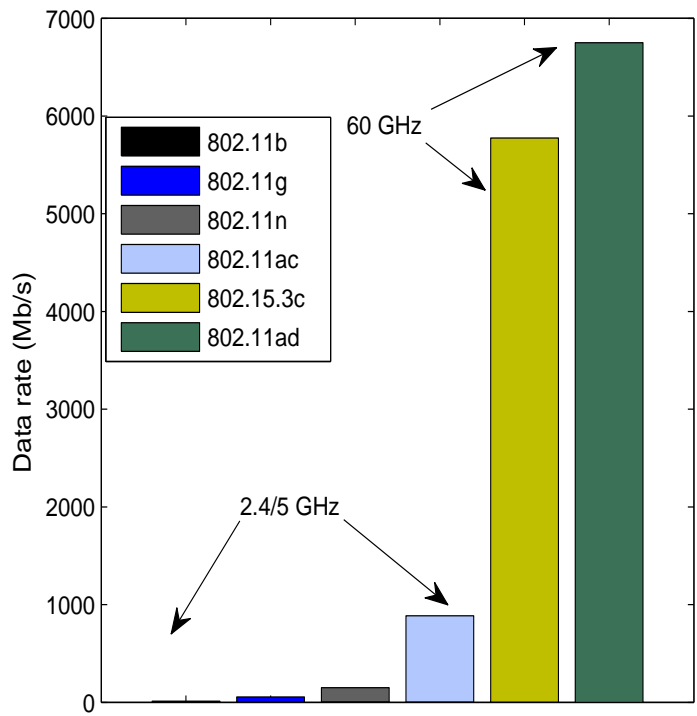}
\label{fig_dataratecompareMAT}}
\caption{Comparison of signal coverage and offered datarates at 2.4/5 and 60\,GHz.}
\label{fig:coverage}
\end{figure}

Fig.~\ref{fig:coverage24} shows the coverage of 2.4\,GHz (left) and 60\,GHz (right) signals in an indoor environment. Radio-wave Propagation Simulator (RPS)~\cite{RPS} employing ray tracing is used to determine the coverage in the indoor area. To calculate the signal power, reflections, up to second order, are considered and all the antennas are assumed to be omnidirectional. The transmission power of antennas is 10\,dBm.  It is clear that three antennas operating at 2.4\,GHz are sufficient to cover the whole area. On the other hand, at 60\,GHz every room needs a dedicated 60\,GHz antenna. This is due to the fact that signal propagation characteristics are significantly different at 2.4\,GHz and 60\,GHz. mmWaves at 60\,GHz do not penetrate through walls. A significant fraction of signal power is absorbed by the walls. This is illustrated by the black ellipses over the blue colored areas in Fig.~\ref{fig:coverage24}. 

Fig.~\ref{fig_dataratecompareMAT} compares the maximum datarates promised by different WLAN standards operating at 2.4/5\,GHz and 60\,GHz frequency bands. Even though IEEE 802.11n and IEEE 802.11ac use very sophisticated PHY layer techniques such as MIMO, MU-MIMO, channel bonding, and frame aggregation at the MAC layer, the expected datarate is much lower compared to what can be achieved at the 60\,GHz frequency band.

It is evident from Fig.~\ref{fig:coverage} that the 2.4\,GHz and 60\,GHz signals complement each other in terms of capacity and coverage. The capacity of 60\,GHz signals is at least ten times higher than the 2.4\,GHz systems. Thus a hybrid solution, involving 2.4\,GHz transmission assisting the 60\,GHz devices can be very effective. Almost every consumer electronic device, such as smartphones, tablets, laptops, cameras, etc., is equipped with WiFi and this trend is expected to continue. Hence, assistance of 2.4/5\,GHz band for 60\,GHz communications seems a pragmatic solution.
\subsection{Hybrid 2.4 and 60\,GHz WLAN Architecture}\label{architecture}
There can be two types of solutions: (i)~utilizing the existing 2.4\,GHz WiFi and IEEE 802.11ad, and modify them accordingly;  or (ii)~a new system other than IEEE 802.11b/g/n and IEEE 802.11ad. The former category is more likely to succeed as majority of wireless communication devices are already equipped with IEEE 802.11b/g/n. 

	\begin{figure*}[]
	\centering
	%\mbox{
	\subfigure[Network architecture.]{ 
	%\centering
	\includegraphics[width=6.0in,height=3.4in]{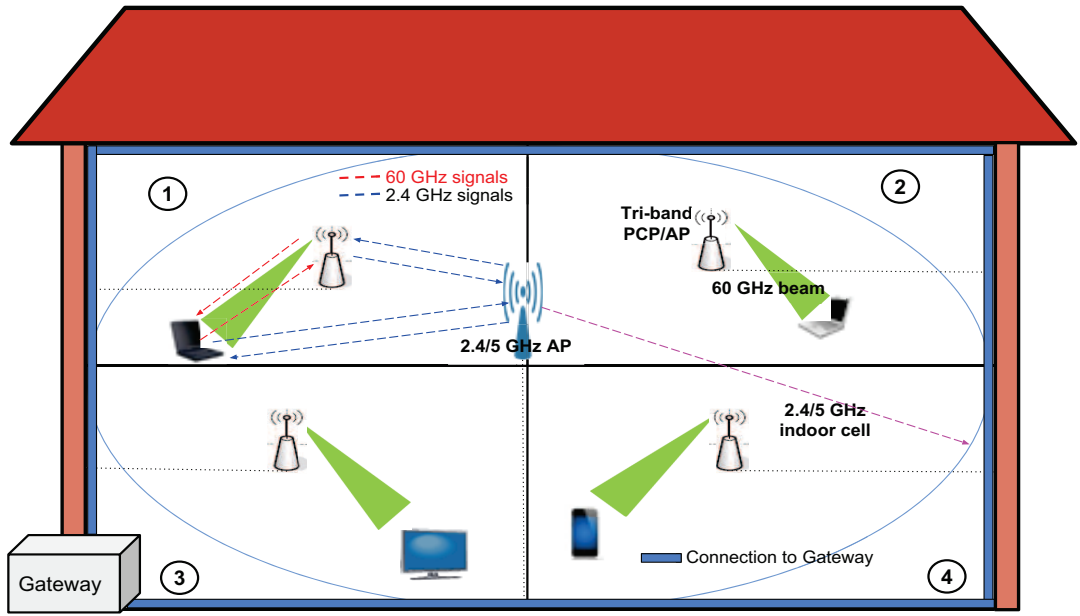}
	\label{5Gindoorscenario}
	}
	\subfigure[Sequence diagram of dual band transmission.]{
	%\centering
	\includegraphics[width=6.0in,height=3.4in]{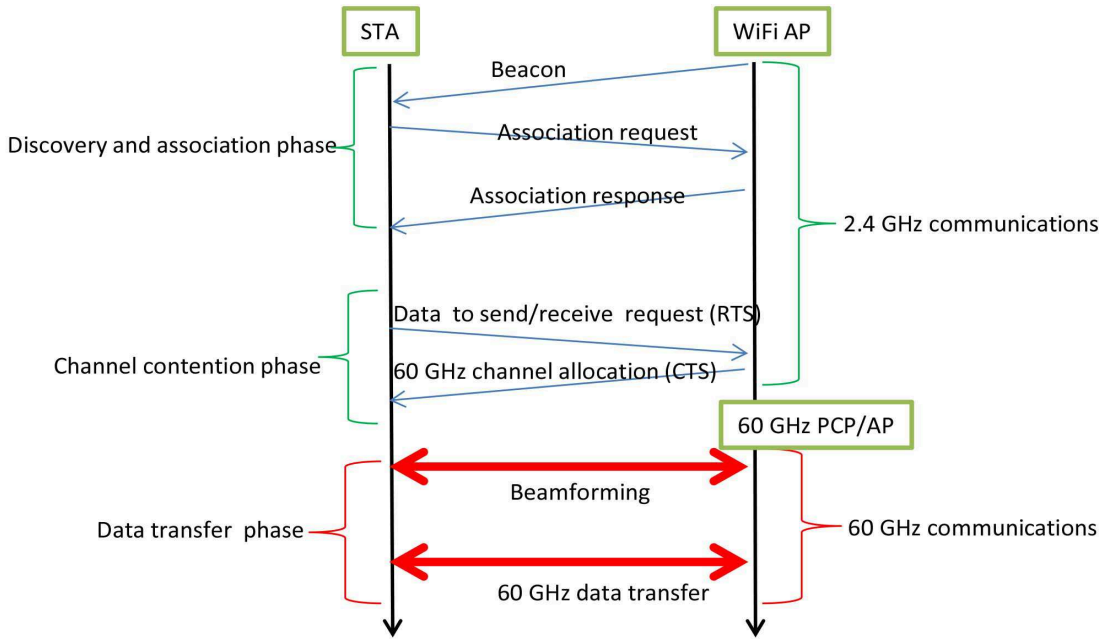}
	\label{fig:dualbandphotoshop}
	}
	%}
	\caption{Interplay of 2.4 and 60\,GHz frequency bands in the proposed indoor network in 5G.}
	\label{fig:5GindoorArchMessage}
	\end{figure*}
One possible approach in the first category could be to use WiFi as a supportive technology to manage the 60\,GHz network. The WiFi AP can cover several 60\,GHz APs and hence, several 60\,GHz APs can be managed by a single WiFi AP. This is the basic idea behind the proposed CogCell architecture. We propose to split control and data plane over 2.4\,GHz and 60\,GHz, respectively. This scheme is similar to the concept of \textit{Phantom Cells}~\cite{phantomcells} proposed for 5G networks.

Fig.~\ref{5Gindoorscenario} shows the conceptual diagram of CogCell architecture. One 2.4\,GHz AP covers all the rooms. Further, every room has a 60\,GHz PCP/AP (802.11ad APs are called PCP/APs) dedicated for high speed data transmission. In a smaller indoor area such as small homes, a single 2.4/5\,GHz AP can be sufficient to provide the coverage but if the indoor area is large (e.g., big office, shopping malls or airports), multiple 2.4/5\,GHz APs would be needed to cover the complete area. Moreover, when areas are separated by walls, they always require a separate 60\,GHz PCP/APs.

In the proposed CogCell architecture, device discovery, association and channel access requests are transmitted over the 2.4\,GHz channel, while data is transmitted over 60\,GHz channel. If a device wants to transmit data, it first sends its request using the 2.4\,GHz frequency band. Thereafter, the appropriate 60\,GHz AP is directed to facilitate the high speed data transmission. IEEE 802.11ad PCP/APs are tri-band devices, hence WiFi AP can communicate with 60\,GHz PCP/AP over 2.4/5 \,GHz. Fig.~\ref{fig:5GindoorArchMessage} shows the schematic of the CogCell architecture.
\subsection{Advantages and Challenges of WiFi and 60\,GHz Interplay}\label{advantage and challenges}
It is to be noted that, other than WiFi, LTE may also assist the mmWaves communications (LTE-WiGig)~\cite{INTEL}. Especially in outdoors, LTE can provide better control functionality instead of WiFi due to its limited range. However, in indoor environments, exploitation of WiFi would be more suitable instead of LTE due to the prevalence of WiFi networks over licensed LTE cells. Furthermore, WiFi would be more suitable for indoor mobility management due to its localization capabilities which are accurate up to a meter and can help in handover between 60\,GHz APs, where room level positioning accuracies would be required. When LTE is used in conjunction with WiGig, the data path could be via LTE base station or there must be different backhaul connectivity to the WiGig AP. In the first case, LTE BS would be the bottleneck and it would defeat the purpose of having WiGig. In the latter case where a backhaul is used for data path via WiGig AP, then it would indeed be similar to CogCell except that LTE handles the control (rather than a WiFi AP as in the CogCell). LTE-WiGig, of course, helps in outdoor environments and it can provide high data rate if backhaul connectivity exists. Now, we briefly describe the advantages and challenges of interplay between WiFi and 60\,GHz.
\subsubsection*{Advantages} The advantages of a hybrid 2.4 and 60\,GHz WLAN system are manifolds: Firstly, isolated (behind the walls) 60\,GHz APs can still facilitate a seamless WLAN experience to the indoor users. Secondly, device discovery and association can be easily performed over 2.4\,GHz. As users move from one room to another room, they are still under the same 2.4\,GHz APs. Thirdly, information sent over the 2.4\,GHz channel can also help in 60\,GHz beamforming procedure. Instead of using two level exhaustive beam searching as in IEEE 802.11ad, devices can estimate the approximate direction of each other using 2.4\,GHz frames. 

Generally, 2.4/5 GHz communications (IEEE 802.11n, IEEE 802.11ac) employ multiple antennas in which approximate direction of arrival can be obtained. Using this rough estimate of direction of arrival, the search space of exhaustive beam searching for 60\,GHz is reduced. A similar approach has been employed in~\cite{BBS_Nitsche} which shows that inferring the direction of 60 GHz transmission using 2.4/5\,GHz can reduce the link setup overhead by avoiding exhaustive beam searching.

\begin{figure*}[t]
\centering
\subfigure[Average device discovery vs number of devices.]{ 
%\centering
\includegraphics[width=4.1in,height=2.8in]{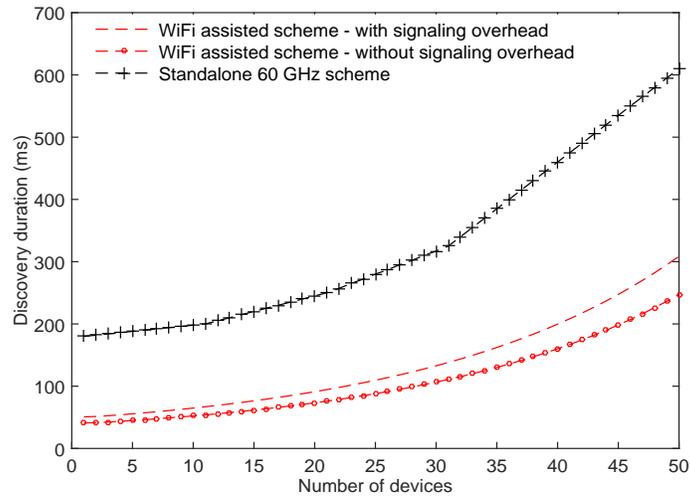}
\label{fig_MDND_eval}
}
%\mbox{
\subfigure[Channel access delay.]{ 
%\centering
\includegraphics[width=3.1in,height=2.8in]{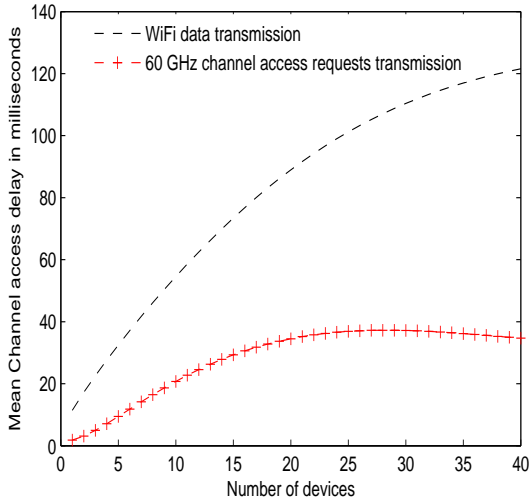}
\label{fig:Channelaccess}
}
\subfigure[Transmission probabilities.]{
%\centering
\includegraphics[width=3.1in,height=2.8in]{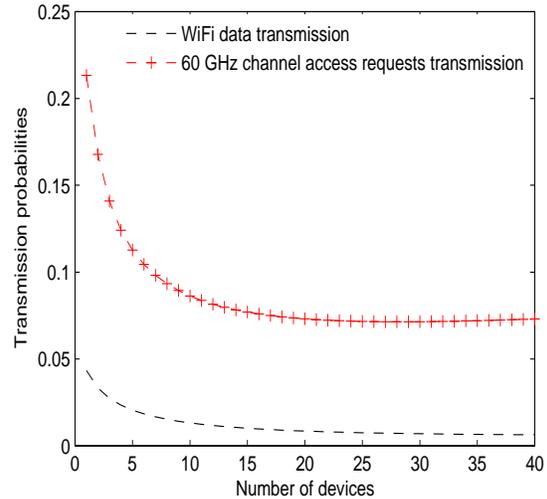}
\label{fig:transmisison}}
%}
\caption{Device discovery time comparison and channel access performance.}
\label{fig:CAPcompare}
\end{figure*}

Fig.~\ref{fig_MDND_eval} shows results from MATLAB simulations for the WiFi assisted device discovery mechanism assuming devices can infer the rough sector estimates using 2.4\,GHz transmissions. The beamwidth of all the devices and PCP/AP is assumed to be 60\degree. All the parameters are listed in Table~\ref{table-hybrid-params}. The results are compared with the standalone 60\,GHz directional device discovery scheme proposed in~\cite{DND_xeuli}. It can be observed that the WiFi assisted scheme is nearly 150\% to 300\% faster than the 60\,GHz directional device discovery scheme. The results also show the effect of signaling overhead due to 2.4\,GHz  control frame transmission which is obtained by including the time required for transmission of extra management frames over 2.4\,GHz. 

Furthermore, the CogCell architecture can reduce the channel access delay because a device can place the data transmission request over 2.4\,GHz channel whenever it wants. On the other hand, in sectorized MAC protocols such as IEEE 802.11ad, a device has to wait for channel access if the 60\,GHz AP is serving a different sector. 

\subsubsection*{Challenges} The hybrid 2.4/5\,GHz and 60\,GHz network poses many challenges also. Firstly, increased number of WiFi devices can hinder the control plane communication. To address this issue, we propose to prioritize the 60\,GHz channel access requests over the 2.4\,GHz requests. We define two categories of frames sent over 2.4\,GHz channel: (i)~60\,GHz channel request frames and (ii)~2.4\,GHz channel request frames by non-60\,GHz devices. We assign different contention window sizes and allowed maximum number of retransmissions for these categories, which are shown in Table~\ref{table-hybrid-params}. Fig.~\ref{fig:Channelaccess} and Fig.~\ref{fig:transmisison} show the MATLAB simulation results for the average channel access delays and transmission probabilities for both type of requests. It can be seen that a significantly faster channel access and higher transmission probabilities can be guaranteed for the 60\,GHz channel requests.
 
Secondly, power consumption of multiple radios working simultaneously can drain the batteries of mobile devices. Hence novel schemes are required to reduce the device power consumption. One possible solution could be to turn on the 60\,GHz radio only when data plane communication is required. Thirdly, 2.4/5\,GHz control plan would also be used as a fall-back options if 60\,GHz data plane is not available. This requires intelligent mechanism to determine when the data plane fall-back should be triggered as 60\,GHz link quality can deteriorate due to multiple reasons such as antenna misalignment due to user movement, blockage due to obstacles, etc. 
 
\begin{table*}[]
\caption{MAC parameters for prioritized control channel access.}
\label{table-hybrid-params}
\centering
%% Some packages, such as MDW tools, offer better commands for making tables
%% than the plain LaTeX2e  
\begin{tabular}{|c|c||c|c|}
\hline
Parameters & Typical values & Parameters & Typical values \\
\hline
Control frame transmission rate & 1\,Mbps & Retry limit[2.4\,GHz] & 5\\
\hline
WiFi datarate & 54\,Mbps & $CW_{max}$[60\,GHz] & 16 \\
\hline
SIFS[2.4\,GHz] & 10$\mu$s & $CW_{max}$[2.4\,GHz] & 256\\
\hline
SIFS[60\,GHz] & 3$\mu$s & RTS & 20 Bytes\\
\hline
Slot time[2.4\,GHz] & 20$\mu$s & CTS & 14 Bytes \\
\hline
Slot time[60\,GHz] & 5$\mu$s & ACK & 14 Bytes\\
\hline
DIFS[60\,GHz] & SIFS + Slot time & PHY Header & 16 Bytes\\
\hline
DIFS[2.4\,GHz] & SIFS + 2$\times$Slot time &MAC Header & 24 Bytes\\
\hline
RIFS & 300$\mu$s & WiFi data & 1024 Bytes \\
\hline
$CW_{min}$[60\,GHz] & 8 & Association request & 1024 Bytes\\
\hline
$CW_{min}$[2.4\,GHz] & 32 & Association response & 16 Bytes\\
\hline
Retry limit[60\,GHz] & 5 & Sector sweep  and feedback frame & 1024 Bytes\\
\hline
\end{tabular}
\end{table*}

\section{Sensor-Assisted Intelligent Beam Switching}\label{beamswitching_sensors}
Communication using narrow beam directional antennas can cause frequent link degradation due to device movement. This is particularly the case with handheld devices such as smartphones, tablets, etc. To set up the directional link between two devices, IEEE 802.11ad  provides a beamforming mechanism for the selection of the best transmit and receive antenna-beam pair. In case of device mobility, beam alignment can be disturbed; this could result in frequent outages of links. If the link quality degrades below a certain limit, the mechanism to select the best beam-pair is restarted (we call this re-beamforming). The re-beamforming procedure involves exhaustive searching in all the possible transmit and receive directions. This leads to a considerable amount of communication overhead as well as degradation of Quality of Service (QoS).

If the next position of the users is known, the PCP/AP and the device can switch their beams to the appropriate beam sectors. We proposes to use the motion sensors such as accelerometer and gyroscope to identify the device movements and predict the next location of device. These sensors are already embedded into most modern devices, hence this method is not unrealistic and is economically viable. To retrieve the useful information from these sensors it is possible to combine the data from two or more sensors. Such combination of sensors is referred to as a virtual sensor. The \textit{rotation vector sensor} is such a virtual sensor, where accelerometer, gyroscope and magnetometer data are fused. The rotation vector sensor gives the orientation of the device relative to the East-North-up coordinates. The azimuth angle from this sensor can be used as an indication of the direction of the user which can assist in identifying the next beam-pairs.

\begin{figure*}[]
\centering
%\mbox{
\subfigure[System diagram of sensor assisted beamforming.]{ 
%\centering
\includegraphics[width=2.8in,height=2.4in]{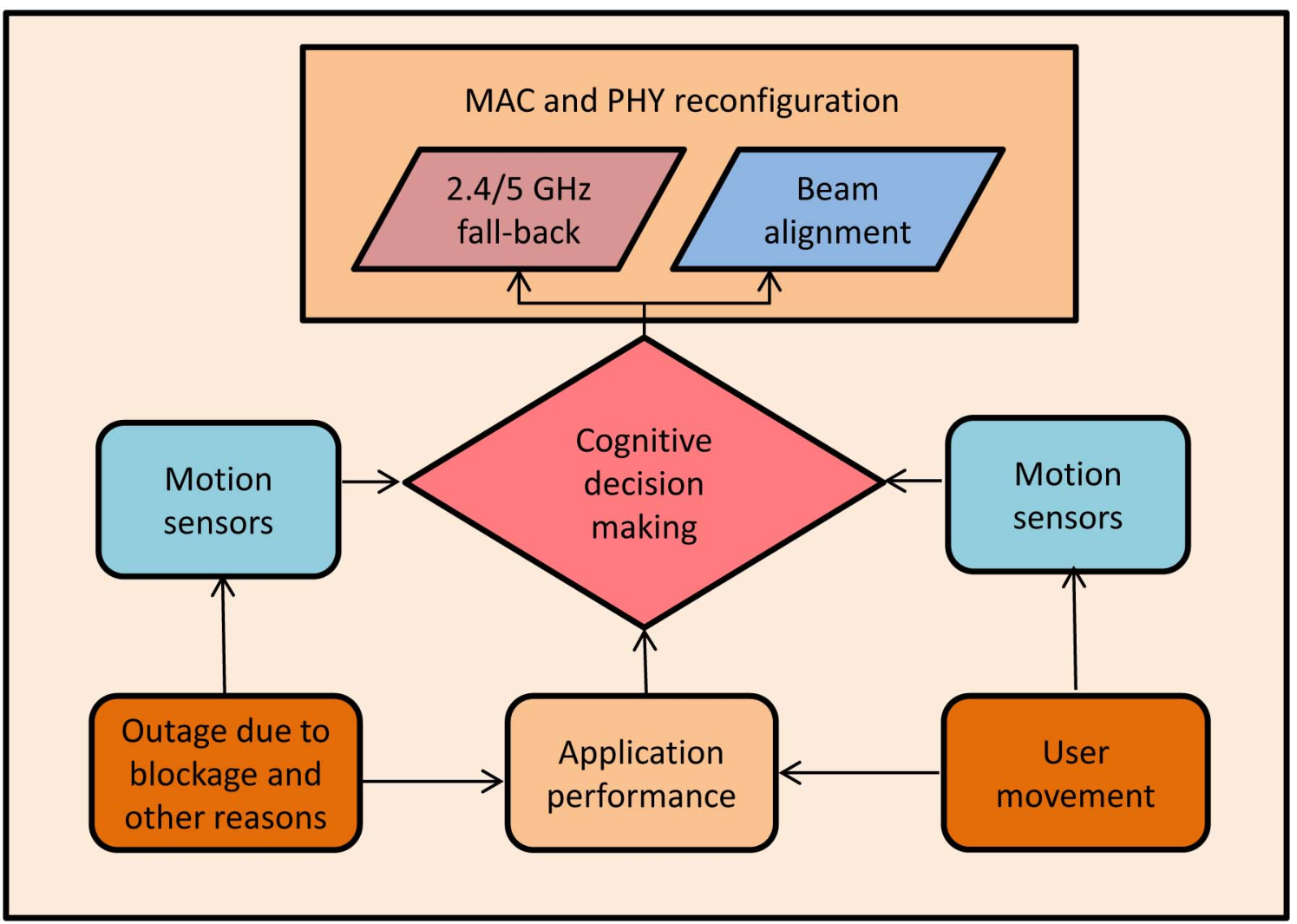}
\label{fig:Systemblockdiagram }
}
\subfigure[Re-beamforming instances.]{
%\centering
\includegraphics[width=2.8in,height=2.4in]{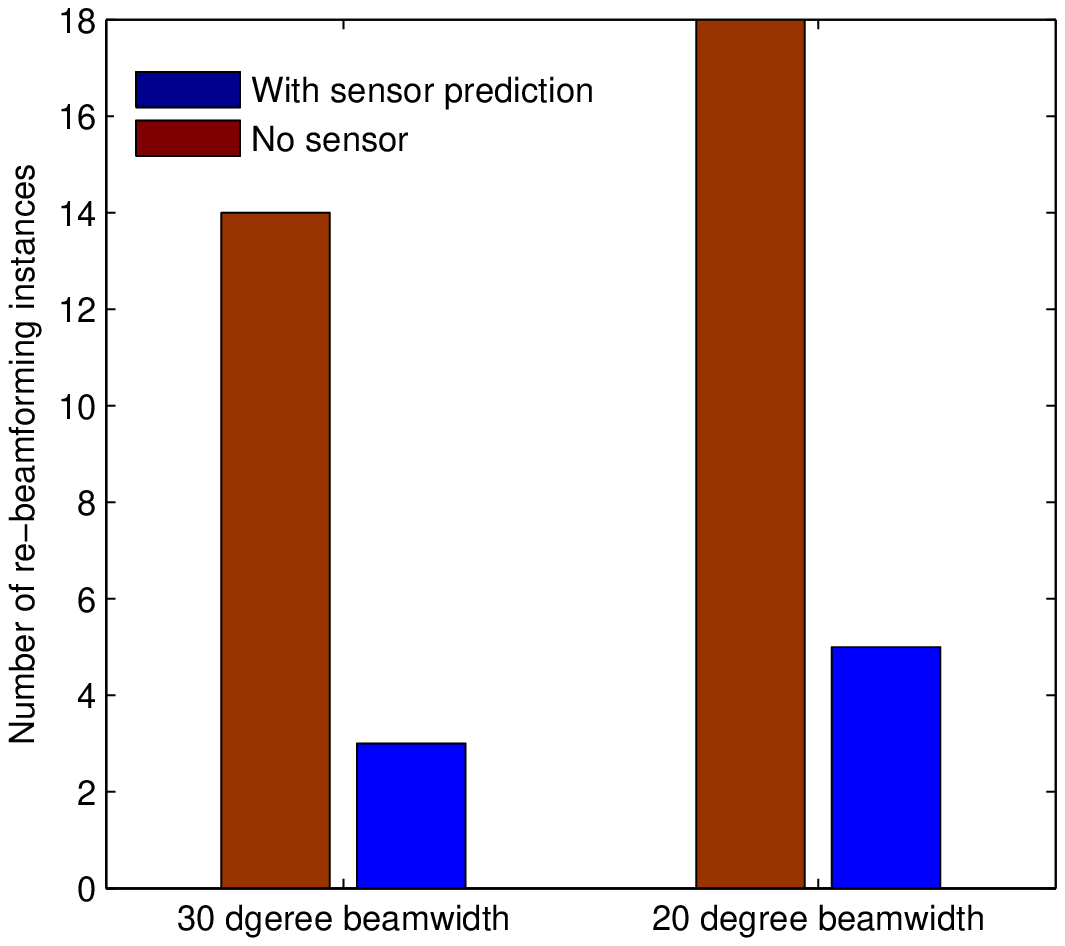}
\label{fig:rebeamfroming_instances}
}
%}
\caption{Sensor-assisted intelligent beamforming.}
\label{fig:routr-and-rebeamfroming}
\end{figure*}

Fig.~\ref{fig:Systemblockdiagram } shows the system diagram of sensor assisted beamforming. Whenever a movement occurs, based on the gathered sensor data, the next location of the user is predicted and beam switching is performed to maintain beam alignment.
Fig.~\ref{fig:rebeamfroming_instances} shows the preliminary simulation results when a device moves along the stated route  in~\cite{ Doff1501:Sensor}. When the PCP/AP beamwidth is 30\degree, 14 instances of re-beamforming are required without using sensor data. On the other hand, with the help of sensor prediction, the number of re-beamformings can be reduced to 4. Similarly, when the PCP/AP beamwidth is 20\degree, instead of 18 re-beamforming instances without sensor prediction, re-beamforming is needed only 5 times using the rotation vector sensor data. 
In this simulation, we assumed that the PCP/AP knows the sensor information. However, in practical scenarios, sensor information needs to be communicated to the PCP/AP. This can be done by including sensor information in the 802.11ad data frames. This preliminary examination of using sensor data for beam switching seems encouraging and requires further investigations.  
\section{Conclusions}\label{conclusion}
In this paper, we proposed a novel indoor network architecture, CogCell, for 5G. The proposed CogCell architecture enables the interplay between 2.4 and 60\,GHz bands for control and data plane transmissions, respectively. CogCell promises a robust multi-Gb/s WLAN experience at 60\,GHz frequency bands enabling faster device discovery and medium access. We believe that the combination of 2.4/5\,GHz WiFi and 60\,GHz communication will play an important role in the indoor networks in 5G era and we showed the approach  to exploit them together. Further, a sensor-assisted intelligent beam switching scheme for 60\,GHz communication was proposed. It was shown that with the help of rotation-vector sensor-data, frequent re-beamforming in the 60\,GHz directional links can be significantly reduced. Thus, link maintainability in 60\,GHz is guaranteed. This results in less requests on WiFi APs leading to efficient use of 60\,GHz and WiFi.
\bibliographystyle{IEEEtran}
\bibliography{references_bibliography}
\section*{BIOGRAPHIES}
\subsection*{Kishor Chandra}
{Kishor Chandra (K.Chandra@tudelft.nl) is currently pursuing his Ph.D. in the Embedded Software group of Delft University
of Technology, The Netherlands. He received his M.Tech. degree in Signal Processing from Indian Institute of Technology, Guwahati, India in 2009 and his B.Eng. in Electronics and Communications Engineering for K.E.C. Dwarahat (Kumaon University), Nainital, India in 2007. Prior to joining Ph.D., he was a Research Engineer working on IP Multimedia Subsystems  with Centre for Development of Telematics (CDOT), New Delhi, India. His  research interests are in the area of 60 GHz communications, 5G Networks and millimeter wave Radio-over-Fiber networks.
\subsection*{R. Venkatesha Prasad}
 R. Venkatesha Prasad (R.R.VenkateshaPrasad@tudelft.nl) received his Bachelor’s degree in electronics and communication
 engineering and M.Tech degree in industrial electronics  from the University of Mysore, India, in 1991 and  1994. He received a Ph.D. degree in 2003 from IISc. During  1996 he worked as a consultant and project associate  for THE ERNET Lab of ECE at IISc. While pursuing his  Ph.D. degree, from 1999 to 2003 he also worked as a  consultant for CEDT, IISc, Bangalore for VoIP application  development as part of Nortel Networks sponsored project.  In 2003 he headed a team of engineers at ESQUBE
 Communication Solutions for the development of various  real-time networking applications. From 2005 to 2012, he
 was a senior researcher at the Wireless and Mobile Communications  group, Delft University of Technology, working  on the EU funded projects MAGNET/MAGNET Beyond  and PNP-2008, and guiding graduate students. From  2012 onward, he has been an assistant professor at the  Embedded Software group at Delft University of Technology.  He is an active member of TCCN and IEEE SCC41,
 and a reviewer of many IEEE transactions and Elsevier  journals. He is on the Technical Program Committees of many conferences, including IEEE ICC, IEEE GLOBECOM,  ACM MM, ACM SIGCHI, and others. He is TPC Co-Chair of  the CogNet Workshop in 2007, 2008, and 2009, and TPC  Chair for E2Nets at IEEE ICC ’10. He is also running PerNets  workshop from 2006 with IEEE CCNC. He was the
 tutorial Co-Chair of CCNC 2009 and 2011, and Demo  Chair of IEEE CCNC 2010. He is Secretary of the IEEE  ComSoc Standards Development Board and Associate Editor of Transactions on Emerging Telecommunications
 Technologies.
\subsection*{Bien Quang}
 Bien Quang (quang.bien@gmail.com) received his Ph.D. degree from Delft University of technology in 2014. He received his
B.S. degree and the M.Sc. degree in Electronics and Telecommunications from Hanoi
University of Technology, Vietnam, in 2001 and 2004, respectively. His research interests include billing,
mobility, home networking, performance analysis of various wireless technologies,
e.g., IEEE 802.15.3 and 802.11. 
\subsection*{I.G.M.M. Niemegeers}
I.G.M.M. Niemegeers (I.G.M.M. Niemegeers@tudelft.nl)  received a degree in electrical engineering from the University of
Ghent, Belgium, in 1970. In 1972 he received an M.Sc.E. degree in computer engineering and in 1978 a Ph.D. degree from Purdue University in West Lafayette, Indiana. From 1978 to 1981 he was a designer of packet switching networks at Bell Telephone Mfg. Cy, Antwerp, Belgium. From 1981 to 2002 he was a professor at the Computer Science and Electrical Engineering faculties of the University of Twente, Enschede, The Netherlands. From 1995 to 2001 he was scientific director of the Centre for Telematics
and Information Technology (CTIT) of the University of Twente, a multi-disciplinary research institute on ICT and applications. From May 2002 until his retirement in 2012, he held the chair Wireless and Mobile Communications at Delft University of Technology, where he headed the Telecommunications Department. He is currently a Professor Emeritus of Delft University of Technology. He was involved in many European research projects, including the EU projects MAGNET and MAGNET Beyond on personal
networks, EUROPCOM on UWB emergency networks, and eSENSE and CRUISE on sensor networks. His present research interests are 5G wireless infrastructures, future home networks, ad hoc networks, personal networks, and cognitive networks. He has (co)authored close to 300 scientific publications and has coauthored a book on personal networks.
\end{document}